\documentclass[twocolumn,showpacs,preprintnumbers,amsmath,amssymb]{revtex4}
\usepackage{graphicx}% Include figure files
\usepackage{dcolumn}% Align table columns on decimal point
\usepackage{bm}% bold math

\begin{document}

\preprint{APS/123-QED}

\title{Generation of 1.5-$\mu$m band time-bin entanglement using spontaneous fiber 
four-wave mixing and planar lightwave circuit interferometers}% 

\author{Hiroki Takesue}
 \email{htakesue@will.brl.ntt.co.jp}
 \altaffiliation\\
\affiliation{%
NTT Basic Research Laboratories, NTT Corporation, 3-1 Morinosato Wakamiya, 
Atsugi, Kanagawa, 243-0198, Japan
}%

\author{Kyo Inoue}
\affiliation{
Department of Electrical, Electronics and Information Engineering, Osaka University, 
2-1 Yamadaoka, Suita, Osaka, 565-0871, Japan. % with \\
}%

\date{\today}% It is always \today, today,

\begin{abstract}
This paper reports 1.5-$\mu$m band time-bin entanglement generation. 
We employed a spontaneous four-wave mixing process in a dispersion shifted fiber, 
with which correlated photon pairs with very narrow bandwidths were generated 
efficiently. To observe two-photon interference, we used planar lightwave circuit based 
interferometers that were operated stably without feedback control. As a result, we 
obtained coincidence fringes with 99 \% visibilities after subtracting accidental 
coincidences, and successfully distributed entangled photons over 20-km standard 
single-mode fiber without any deterioration in the quantum correlation.  

\end{abstract}

\maketitle

%%%%%%%%%%%Introduction%%%%%%%%%

In recent years, the generation and distribution of entangled photon pairs have been 
studied intensively with a view to realizing such forms of quantum communication as 
quantum cryptography \cite{ekert,bbm92}, quantum teleportation \cite{bennett3} and 
quantum repeaters \cite{briegel}. Although practical entanglement sources based on 
parametric down conversion (PDC) have been reported and widely used in the short 
wavelength band \cite{kwiat1,kwiat2}, what is needed most for scalable quantum 
communication networks over optical fiber is a practical entanglement source in the 
1.5-$\mu$m band, where silica fiber has its minimum loss. 
Several polarization entanglement sources in the 1.5-$\mu$m band have already been 
reported \cite{yoshizawa,takesue,li,takesue2}. However, when transmitting 
polarization entangled photons over optical fiber, polarization mode dispersion (PMD) 
causes decoherence, which limits the transmission length. Therefore, polarization 
entanglement is not the best choice for quantum communication over optical fiber. 

Time-bin entanglement has been proposed to overcome this problem \cite{brendel}. 
This scheme is based on qubits spanned by two time slots instead of two polarization 
modes, and so is unaffected by PMD. Although degenerated photon pairs in the 
1.3-$\mu$m band \cite{brendel} and nondegenerated photon pairs in the 
1.3/1.5-$\mu$m band \cite{marc} have been reported with this scheme, the use of 
photon pairs where both photons are in the 1.5-$\mu$m band is obviously the most 
effective way of increasing the transmission length. Recently, a Hong-Ou-Mandel 
experiment using quantum correlated photon pairs both in the 1.5-$\mu$m band was 
demonstrated by Halder et al. \cite{halder}. However, the direct observation of the 
degree of entanglement on time-bin entangled photons in the 1.5-$\mu$m band has yet 
to be achieved.
The large bandwidths of the previous entangled photon-pair sources pose another 
problem. Although there have been a few reports on entangled photon-pair sources 
with narrow bandwidths \cite{li,halder}, most of the previous time-bin entanglement 
experiments employed PDC that generated photon pairs with a typical bandwidth of 
$\sim$10 nm \cite{marc}. This relatively large bandwidth made it difficult to 
distribute such photons over a standard single-mode fiber (SMF) with a zero dispersion 
wavelength of $\sim$1.3 $\mu$m, because of the pulse broadening caused by the 
large chromatic dispersion of the SMF. Since most installed fibers are standard SMF, a 
long-distance transmission capability over a standard SMF is an important requirement 
for the entangled photons for quantum communication utilizing existing fiber 
networks. 
Also, fiber interferometers have been used in previous experiments to observe 
two-photon interference. Such interferometers are unstable and need feedback control 
to ensure their long-term stability \cite{marc}. 

In this paper, we report a time-bin entanglement generation experiment, in which the 
degree of entanglement was directly measured for time-bin entangled photon pairs 
both in the 1.5-$\mu$m band. We employed the spontaneous four-wave mixing 
(SFWM) process in dispersion-shifted fiber (DSF) \cite{takesue,li}, with which we 
obtained quantum correlated photon pairs with narrow bandwidths. To observe 
two-photon interference, we used 1-bit delay interferometers made of planar lightwave 
circuits (PLC), which are silica-based optical waveguides fabricated on silicon 
substrates \cite{honjo}. The excellent stability of the PLC interferometers enabled 
long-term measurements to be performed without feedback control. As a result, we 
observed a coincidence fringe with 99 \% visibility after subtracting accidental 
coincidences. Moreover, we successfully distributed the entangled photons over 20 km 
of standard SMF without any deterioration in the quantum correlation, thanks to the 
narrow bandwidths of the entangled photons.

%%%%%%%%%%%experimental setup%%%%%%%%%%

\begin{figure}[htb]

\centerline{\includegraphics[width=\linewidth]{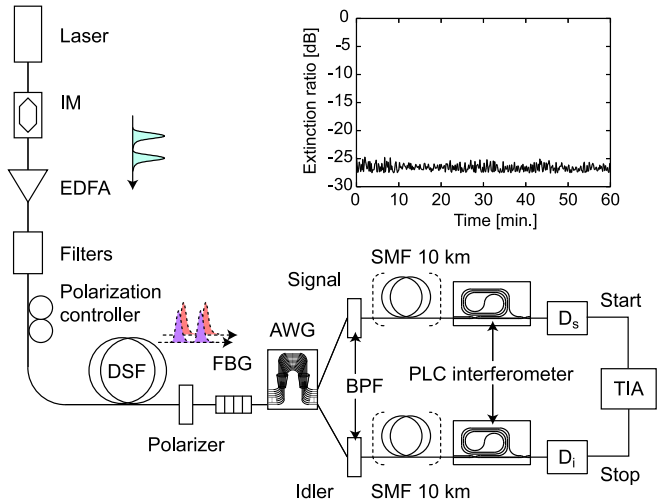}}
\caption{Experimental setup and extinction ratio of PLC interferometer monitored for 
an hour.}
\label{1}

\end{figure}

Figure \ref{1} shows the experimental setup. A continuous lightwave with a 
wavelength of 1551.11 nm from an external cavity semiconductor laser is modulated 
into double pulses with a 1-ns interval using an intensity modulator (IM). The pulse 
width and repetition frequency are 90 ps and 100 MHz, respectively. The coherence 
time of the laser output is $\sim$10 $\mu$s, which is far larger than the temporal 
interval between the pulses \cite{coh}. The double pulses are amplified by an 
erbium-doped fiber amplifier (EDFA), and filtered to reduce amplified spontaneous 
emission noise from the EDFA. The pulses are then input into a 2.5-km DSF with a 
zero-dispersion wavelength of 1551 nm after polarization adjustment. 
In the DSF, time-correlated photon pairs are generated through SFWM, with the 
double pulses as the pump. The pump, signal and idler frequencies, $f_p$, $f_s$ and 
idler $f_i$, respectively, have the following relationship: $2 f_p = f_s + f_i$. 
We set the pump power at a relatively small value so that the probability of both pulses 
generating photon pairs becomes very low. As a result, we can generate a time-bin 
entanglement, which is a superposition of two-photon states in different time instances 
expressed as 
\begin{equation}
|\Phi\rangle = \frac{1}{\sqrt{2}} \left(|1\rangle_s |1\rangle_i + e^{i \phi} 
|2\rangle_s |2\rangle_i \right). \label{2timebin}
\end{equation}
The expression $|k\rangle_x$ represents a state in which there is a photon in the $k$th 
time slot in a mode $x$, signal $(s)$ or idler $(i)$. 
$\phi$ is a relative phase term that is equal to $2 \phi_p$, where $\phi_p$ is the 
phase difference between the two pump pulses. 
With pump degeneration, the efficiency of four-wave mixing in a long fiber reaches 
maximum when the pump, signal and idler photons are all in the same polarization 
state \cite{kyo}, so the correlated photons generated in the SFWM are expected to 
have the same polarization state as the pump photons.  
The phase matching condition of the SFWM is satisfied by setting the pump 
wavelength at the zero-dispersion wavelength of the DSF, as in our experiment. 
The use of longer fiber increases the efficiency of the SFWM process, by which we 
can increase photon pair production within the fixed bandwidths determined by the 
filters for separating signal and idler photons. 

The output light from the DSF passes through a polarizer, whose function is explained 
later, and is input into fiber-Bragg gratings (FBG) to suppress the residual pump 
photons. Then, the photons are input into an arrayed waveguide grating (AWG), which 
separates the signal and idler photons. AWG output channels with peak frequencies of 
$-$400 and $+$400 GHz from the pump photon frequency are used for the signal and 
idler, respectively. 
The output photons from the AWG are filtered using dielectric bandpass filters (BPF) 
to further suppress the pump photons. With the FBGs, AWG and BPF, the 
transmittance of the pump photons becomes $<$-125 dB smaller than those of the 
signal and idler photons. 
The 3-dB bandwidths of the signal and idler channels are both 25 GHz ($\simeq 
0.2$ nm), which is very small compared with most previous photon-pair sources, and 
approximately 1/3 and 1/4 of the bandwidths of the narrowband photon-pair sources 
reported in \cite{li} and \cite{halder}, respectively.

The signal and idler photons are then input into PLC interferometers, which convert a 
state $|k\rangle_x$ to
$\left(|k\rangle_x + e^{i\theta_x} |k+1\rangle_x \right)/\sqrt{2}$, where 
$\theta_x$ is the phase difference between the two paths of the interferometer for 
mode $x$, and can be tuned by changing the temperature. 
Then, after passing through the PLC interferometers, the state shown by Eq. 
(\ref{2timebin}) becomes
\begin{eqnarray}
|\Phi\rangle &\to&
|1\rangle_s |1\rangle_i +  (e^{i(\theta_s +\theta_i)} +e^{i\phi}) |2\rangle_s 
|2\rangle_i \nonumber \\
& & + e^{i(\phi+\theta_s +\theta_i)} |3\rangle_s |3\rangle_i, \label{after}
\end{eqnarray}
where an amplitude term that is common to all product states is omitted for simplicity 
and non-coincident terms are discarded because they do not appear in a coincidence 
measurement. Thus, we can observe two-photon interference at the second time slot.  
The photons output from the PLC interferometers are input into photon detectors based 
on InGaAs avalanche photodiodes (APD) operated in a gated mode with a 4-MHz 
frequency \cite{memo}. By applying a 1-ns-wide gate only in the second time slots, 
we post-select the second term on the right hand side of Eq. (\ref{after}). 
The quantum efficiencies of the detectors for the signal ($D_s$) and idler ($D_i$) were 
8.0 and 9.5 \%, respectively. The dark count rate per gate was 4 $\times 10^{-5}$ for 
$D_s$ and 7.5 $\times 10^{-5}$ for $D_i$. Both PLC interferometers had a 2.0-dB 
loss. The signal and idler arm losses were 7.0 and 7.1 dB, respectively. 
The detection signals from $D_s$ and $D_i$ are input into a time-interval analyzer 
(TIA) as start and stop pulses, respectively, for the coincidence measurement.

In previous reports on SFWM, noise photons possibly due to spontaneous Raman 
scattering (SRS) were observed \cite{li2,kyo2}, which degraded the visibility of the 
two-photon interference \cite{takesue,li}. SFWM photons have the same polarization 
as the pump, while SRS photons are depolarized. Therefore, we can suppress the cross 
polarization component of SRS photons by placing a polarizer after the DSF as in Fig. 
\ref{1} \cite{li2}. We adjusted the polarization of the pump photons with a 
polarization controller, so that we could maximize the ratio of the SFWM photon 
number to the SRS photon number. 

%%%%%%%%%%correlation measurement%%%%%%%%

\begin{figure}[htb]

\centerline{\includegraphics[width=.8\linewidth]{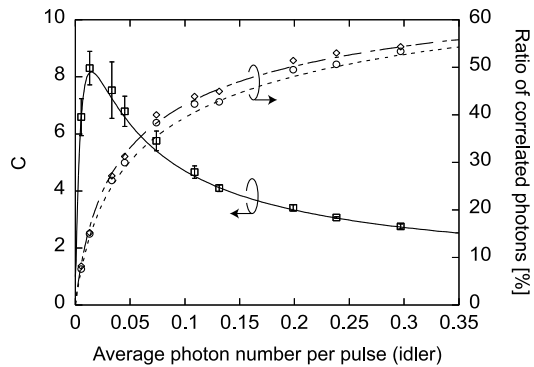}}

\caption{$C$ value and ratio of correlated photons as a function of average idler 
photon number. Fitted curves were produced based on the pump-power dependence 
functions of the number of correlated and noise photons obtained in the experiment.}
\label{c}
\end{figure}

We first measured the time-correlation of photon pairs without inserting PLC 
interferometers. We measured the coincidence rates at matched and un-matched slots, 
which we refer to as $R_m$ and $R_{um}$, respectively. A coincidence in a matched 
slot is a ``true" coincidence caused by photons generated with the same pump pulse. A 
coincidence in an un-matched slot corresponds to an accidental coincidence caused by 
photons generated by different pump pulses and it appears at different time instances 
separated by the detector gate period. Therefore, the ratio $C=R_m/R_{um}$ is a good 
figure of merit: $C>1$ implies the existence of a time correlation. In our TIA 
measurement, $C$ can be expressed as
\begin{equation}
C = \frac{R_m}{R_{um}} = \frac{\mu_c \alpha_s \alpha_i}{c_s c_i} +1, 
\label{cc}
\end{equation}
where $\mu_c$, $\alpha_s$ and $\alpha_i$ show the average number of correlated 
photon pairs per pulse, the system transmittance (including detector quantum 
efficiency) for the signal, and that for the idler, respectively. $c_s$ and $c_i$ are the 
average count rate per pulse for the signal and idler, respectively, and can be expressed 
as 
\begin{equation}
c_x = (\mu_c + \mu_{xn}) \alpha_x + d_x \label{i}
\end{equation}
where $x=s,i$.
Here, $\mu_{xn}$ and $d_x$ are the average number of noise photons per pulse in 
mode $x$ and the dark count rate for $D_x$. 
Using Eqs. (\ref{cc}) and (\ref{i}), the experimentally obtained $C$, the system 
transmittances and the count rates, we can estimate the ratio between correlated and 
noise photons. We changed the average photon number per pulse by changing the 
pump power, and measured the coincidence rate. The squares in Fig. \ref{c} show 
$C$ as a function of the average number of idler photons per pulse $\mu_i$($=\mu_c 
+ \mu_{in}$). $C$ reached 8.3 at an average idler photon number of 0.01. The circles 
and diamonds show the ratio of correlated photons for the signal and idler, respectively, 
which were calculated using Eq. (\ref{cc}). The signal was slightly noisier than the 
idler because of the slight difference in the SRS gain for the Stokes and anti-Stokes 
side. Although we can improve the portion of correlated photons as we increase the 
average number of photons, $C$ degrades because the intrinsic accidental 
coincidences caused by the Poissonian statistics of the SFWM photons pairs increase. 
In the following two-photon interference experiment, we adopted $\mu_i=0.13$, 
where the portion of correlated photons was $\sim$45 \% and $C \simeq 4$. 

%%%%%%%%%%extinction ratio%%%%%%%%%%%%

We then measured the long-term stability of a PLC interferometer and the pump laser 
frequency. 
A continuous lightwave from the pump laser was input into a PLC interferometer. We 
adjusted the phase difference induced in the two paths so that we could observe a dark 
fringe from one of the output ports. We monitored the extinction ratio, which is the 
ratio of the powers from the two output ports, for one hour. The result is shown in the 
top right in Fig. \ref{1}. The extinction ratio was better than -25 dB throughout the 
measurement. This means that we could operate our PLC interferometer and the pump 
laser stably for at least one hour without feedback control.

%%%%%%%%%%coincidence fringes%%%%%%%%%%
\begin{figure}[th]
\centerline{\includegraphics[width=.8\linewidth]{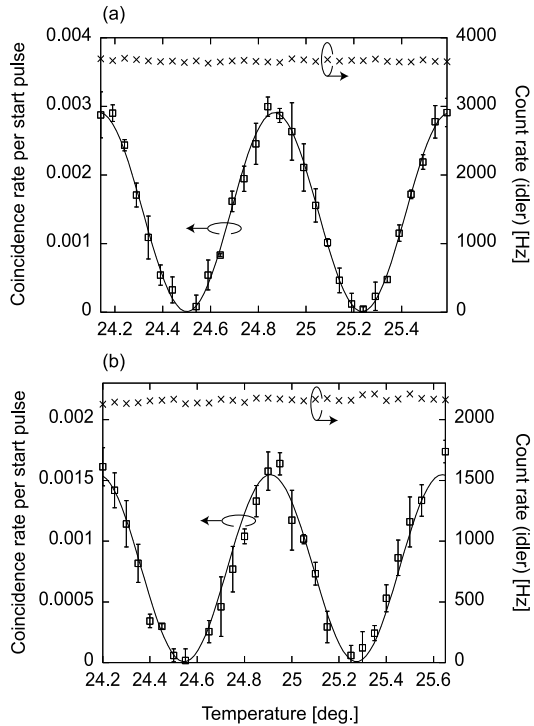}}

\caption{(a) Coincidence rate per start pulse and count rate (idler) as a function of the 
temperature of the PLC interferometer for the idler. (b) After 10 km x 2 SMF 
transmission.}
\label{fringe}
\end{figure}

Next, we undertook a two-photon interference experiment. We fixed the temperature of 
the PLC interferometer for the signal, changed that for the idler, and measured the 
coincidence. The peak power of the pump pulse was $\sim$80 mW and the average 
number of photons per pulse for both signal and idler was set at $\sim$ 0.13. The 
coincidence rate per start pulse is shown by the squares in Fig. \ref{fringe} (a), where 
accidental coincidences are subtracted. The count rate of $D_s$ and the peak 
coincidence rate were $\sim$2800 and 4.5 Hz, respectively. While the count rate of 
$D_i$, shown by the x symbols in Fig. \ref{fringe} (a), remained unchanged, we 
observed a deep modulation of the coincidence rate as we changed the temperature. 
The visibility of the fitted curve for the coincidence rate was as large as 99.3 \%. 
When the accidental coincidences were included, the visibility was 61.4 \%. The high 
extinction ratio and stability of the PLC interferometers contributed to this good 
visibility. 

Finally, we inserted a 10-km standard SMF with a 1310-nm zero-dispersion 
wavelength in each path, and again measured the two-photon interference. The result is 
shown in Fig. \ref{fringe} (b). Although the coincidence rate decreased due to the 
additional fiber loss of $\sim$2.3 dB for each path, the observed visibility remained at 
99.0 \% with the accidental coincidences subtracted. The visibility was 60.8 \% with 
the accidental coincidences included. This result means that we observed no 
decoherence caused by SMF chromatic dispersion, and so the entangled photon source 
based on fiber SFWM is suitable for fiber transmission because it can generate photon 
pairs with narrow bandwidths.

%%%%%%%%%%%%%%Discussion%%%%%%%%%%%%

%%%%%%Conclusion%%%%%

In conclusion, we described the generation and observation of time-bin entanglement 
in the 1.5-$\mu$m band. SFWM in a DSF was used for the efficient generation of 
quantum-correlated photon pairs with narrow bandwidths. We used PLC 
interferometers, whose high extinction ratio and good stability made it possible to 
observe two-photon interference with a very high visibility of 99 \%. 
A 20-km (10 km $\times$ 2) SMF transmission experiment was also undertaken and 
no degradation in visibility was observed thanks to the narrow bandwidths of the 
photon pairs. The excellent stability and long-distance transmission capability over 
standard SMF means that we can expect the above method to provide a practical 
entanglement source for quantum communication over optical fiber.

This work was supported in part by National Institute of Information and 
Communications Technology (NICT) of Japan.


\begin{thebibliography}{99}
\bibitem{coh} The relative phase $\phi$ in Eq. (\ref{2timebin}) is equal to $2 \Delta 
\phi_p$, where $\Delta \phi_p$ denotes the phase difference of the pump light 
between the first and second time slots \cite{takesue}. This means that, if the 
coherence time of the pump is shorter than the temporal interval of these two slots, the 
relative phase $\phi$ is undetermined. Thus, the coherence time of the pump must be 
larger than the pulse interval in order to obtain an entangled pure state such as Eq. 
(\ref{2timebin}).
\bibitem{memo} An InGaAs APD for the long-wavelength band generally has a larger 
dark count probability than that of a Si-APD for the short-wavelength band. Therefore, 
to reduce noise counts, an InGaAs APD is normally biased above its breakdown 
voltage for a short period of time, which is called a gated mode operation. The upper 
limit of the gate frequency, which was 4 MHz in our experiment, is determined by the 
after-pulse characteristics of an APD.
\end{thebibliography}
\end{document}